\documentclass[]{article}
\usepackage{todonotes}
\usepackage{microtype}
\usepackage{url}
\usepackage{mathtools}
\usepackage{amsmath}
\usepackage{pgfplotstable}
\usepackage{algorithm}
\usepackage[noend]{algpseudocode}
\usepackage{amssymb }
\usepackage[detect-all]{siunitx}
\usepackage{booktabs}

\newcommand\RepackingStations{\ensuremath{S_{\text{participating}}}}
\newcommand\NonParticipatingStations{\ensuremath{S_{\text{non-participating}}}}

\newcommand\Winners{\ensuremath{S_{\text{winners}}}}

\title{Assessing Economic Outcomes in Simulated Reverse Clock Auctions for Radio Spectrum}
\author{Neil Newman, Kevin Leyton-Brown, Paul Milgrom, Ilya Segal}
\def\UHF{UHF\,}
\def\VHF{VHF\,}

\begin{document}

\maketitle

\begin{abstract}
We investigate the economic outcomes that result under simulated bidder behavior in a model of the FCC's reverse auction for radio spectrum. In our simulations, limiting our notion of efficiency to the reverse auction in isolation, the reverse clock auction achieves very efficient solutions, the FCC's scoring rule greatly reduces the total payments to TV broadcasters at the cost of some efficiency, and using a poor feasibility checker can have grave consequences both in terms of the auction's cost and efficiency.
\end{abstract}

\section{Introduction}\label{sec:Intro}
Over 13 months in 2016--17, the Federal Communications Commission (FCC) ran a novel spectrum auction referred to as the \emph{incentive auction}. What made the incentive auction unique compared to prior spectrum auctions is that rather than allocating unlicensed spectrum, this auction reallocated spectrum held by television broadcasters. 

The incentive auction is interesting to study both due to its novelty and its economic importance: the resale of spectrum yielded \$19.8 billion, and the spectrum reallocation is expected to improve social welfare by putting spectrum to a higher-value use.

The incentive auction is composed of two related parts: a \emph{reverse auction} in which the FCC acquires spectrum from some broadcasters while remaining broadcasters are ``repacked'' into a reduced set of channels, and a \emph{forward auction} in which the FCC sells the cleared spectrum to mobile carriers. 

The reverse auction needs to decide which stations to acquire and which stations to reassign to a new channel in the reduced band. Reassignment is not always possible, since stations broadcasting on nearby channels can interfere with each other, making certain sets of stations impossible to repack.  One sensible efficiency-related goal for the auction might be, for any given clearing target, to maximize the total value of the stations that remain on the air, or equivalently, to minimize the total value of the stations that the FCC purchases. A second efficiency-related goal is to clear as many channels as possible within a limited budget.

In practice, optimizing for the first goal is computationally intractable. Reasoning about whether a given set of stations can be packed into a set of channels is an NP-complete problem on its own, and in practice the more challenging problem of solving for the packing which minimizes the value of winning stations cannot be done in a timely manner at the national scale of the auction. For our simulations, we assume that the set of channels to be cleared is fixed. Given that assumption, we use the terms ``efficiency'' and "cost" below as abbreviations that refer to measures of the total value of the stations that remain on air and the total payments made to broadcasters that go off air. In the context of the incentive auction, both are measures of efficiency-related performance goals.

Given that maximizing the value of stations that remain on air is impractical, the FCC is instead using a descending-clock auction in which, implicitly, stations are approached in a round-robin fashion and made a series of decreasing price offers for their spectrum as long as they remain packable (pass a \emph{feasibility check}) in the set of available channels. If a station declines the initial or any subsequent offer, it permanently exits the auction and is guaranteed to be packed. As the band fills up, a station that has yet to exit may cease to be packable: this can happen either because the feasibility checker can prove that the station cannot be packed, or because a feasible channel assignment was not found in the amount of time the auction is able to tolerate. Either way, such a station is \emph{frozen} and gets paid the amount to which it most recently agreed.

Do clock auctions perform well in this application? On the question of incentives, the answer is ``yes'': Clock auctions are ``truthful'' mechanisms: no bidder can increase its payoff in a clock auction by pretending to have a station value different from its true value. According to Li \cite{li2015obviously}, these auctions have the even stronger property of being \emph{obviously strategy-proof}. In this context, that means that a bidder can infer the optimality of reporting truthfully even without understanding the details of how its clock prices are set. While that is an important advantage, it does not answer the quantitative algorithmic question: If participants bid truthfully, how close will the clock auction design bring us to the original goal of minimizing the value of winning stations? The answer is not straightforward: it depends on more specific design decisions, including how feasibility checking is performed and the order in which stations exit the auction, which in turn depends on the relative clock prices during the auction, which are parameterized by the \emph{scoring rule}. In addition, performance depends
on how much spectrum is cleared and the values stations place on their broadcast rights. These decisions and parameters impact not only the efficiency of the auction, but also its cost. For example, a feasibility checker that is unable to find feasible assignments when they exist might cause a station to be bought unnecessarily, reducing efficiency, or else to be purchased too early when it could have been bought more cheaply later in the auction. 

The purpose of this paper is to determine how well a descending clock auction can approximate the optimal packing and to examine the impact of the feasibility checker and scoring rule on the efficiency and cost of the reverse auction. Our primary investigative tool is simulation: we instantiated various parameterizations of a reverse auction simulator using different feasibility checkers and scoring rules. We repeated our simulations using different station valuations to test the robustness of our results. We computed the solution and cost of running a  Vickrey--Clarke--Groves (VCG) mechanism, which implements the efficient solution, and compared our simulations against these outcomes.

The structure of the rest of the document is as follows: Section~\ref{sec:FeasibilityProblem} formalizes station packing, Section~\ref{sec:ComparingOutcomes} explains how we compare simulation results, Section~\ref{sec:VCG} describes the VCG mechanism in this setting, Section~\ref{sec:ReverseAuction} describes the reverse auction mechanism and our simulator, and Section~\ref{sec:Results} details our experiments and findings.

\section{Station packing}\label{sec:FeasibilityProblem}

Each television station in the US and Canada $s \in \mathcal{S}$ was assigned to a channel $c_s \in \mathcal{C} \subseteq \mathbb{N}$ prior to the auction that ensured that it would not excessively interfere with other, nearby stations. The FCC reasons about what interference would be harmful via a complex, grid-based physical simulation (``OET-69'' \cite{oet69}), but has also processed the results of this simulation to obtain a Constraint Satisfaction Problem (CSP) style formulation listing forbidden pairs of stations and channels, which it has publicly released \cite{constraintfiles}. Let $\mathcal{I} \subseteq (\mathcal{S}\times\mathcal{C})^2$ denote a set of \emph{forbidden station--channel pairs} $\{(s,c),(s',c')\}$, each representing the proposition that stations $s$ and $s'$ may not concurrently be assigned to channels $c$ and $c'$, respectively. The effect of the auction is to remove some broadcasters from the airwaves completely, and to reassign channels to the remaining stations from a reduced set of channels. This reduced set is defined by a \emph{clearing target}: some channel $\overline{c} \in \mathcal{C}$ such that all stations are only eligible to be assigned channels from $\overline{\mathcal{C}} = \{c \in \mathcal{C} \mid c < \overline{c}\}$. The clearing target is fixed for each stage of the reverse auction. Each station can only be assigned a channel on a subset of $\overline{\mathcal{C}}$, given by a \emph{domain} function $\mathcal{D}: \mathcal{S} \to 2^{\overline{\mathcal{C}}}$ that maps from stations to these reduced sets. A feasible packing $\gamma : S \to \overline{C}$  is one that assigns each station a channel from its domain that satisfies the interference constraints: i.e., for which $\gamma(s) \in D(s)$ for all $s\in S$, and $\gamma(s) = c \Rightarrow \gamma(s') \not= c'$ for all $\{(s,c),(s',c')\} \in I$. 


In this report, we limit attention to \UHF{} stations and channels (the real auction also interacts with the \VHF{} band). This implies that there are only two possible outcomes for a station: it remains on-the-air or goes off-air.

\section{Comparing Reverse Auction Outcomes}\label{sec:ComparingOutcomes}

The outcome of the reverse auction is a channel assignment $\gamma$ for stations that remain on-the-air and a set of prices that winning stations will be paid to go off-air. We compare reverse auction outcomes based on their efficiency and cost. A natural way to measure efficiency is by the value preserved by the auction, in other words by defining an efficient repacking as one that maximizes the total value of the participating stations that remain on-the-air. We instead choose to measure efficiency by the total value \emph{lost} by the auction, defining an efficient repacking as one that minimizes the total value loss of winning stations, which for any given station is measured as the difference between that station's value for broadcasting in its home band and its post-auction band. Note that an efficient repacking satisfying one definition also satisfies the other. We prefer the latter definition because it is only influenced by stations that a feasibility checker is unable to repack in their home bands and therefore does not assign credit for easy to repack stations. For example, holding everything else constant, increasing the value of an easy to repack station arbitrarily high would make the value preserved arbitrarily high as well, but value loss would be unaffected. We therefore feel that the value loss definition allows for more meaningful comparisons between different feasibility checkers. If we can find an efficient repacking $\gamma^*$, then we have an upper bound on efficiency. We can use $\gamma^*$ to relate the total value loss of another repacking to the optimal total value loss through a value loss ratio
\begin{equation}
\text{Value Loss Ratio} = \frac{\sum_{s \not\in \gamma^*} \mathcal{V}\left(s\right)}{\sum_{s \not\in \gamma} \mathcal{V}\left(s\right)}.
\end{equation}
Since $\gamma^*$ is optimal, a value loss ratio will always be weakly greater than 1. 

For computing cost, we define a price function $\mathcal{P}: S \to \mathbb{R}$ that maps from a station to its payment. Then an outcome's cost is
\begin{equation}
\text{Cost} = \sum_{s \in \Winners{}} \mathcal{P}\left(s\right).
\end{equation}
We prefer outcomes with value loss ratios close to 1 and low cost.  It is straightforward to compare two outcomes if one Pareto dominates the other with respect to efficiency and cost, otherwise any comparison is at least partially subjective.

\section{VCG and Optimal packing}\label{sec:VCG}

It has long been known that when bidders can have any values, the VCG auction is the only truthful mechanism that always selects efficient outcomes and entails no payments to losing bidders \cite{green1979incentives}. We now specialize the VCG auction to code for the relevant constraints and to define the benchmark we will use to evaluate the performance of the FCC's alternative auction mechanism.

A station's on-air value is \[ \begin{cases} 
\mathcal{V}\left(s\right) & s\text{ remains on air;} \\
0 & \text{otherwise}.
\end{cases}
\]

We note that stations are divided into two sets: participating stations \RepackingStations{} and non-participating stations \NonParticipatingStations{}. Importantly, if a station does not participate, it must remain on air. We create a Boolean variable $x_{s,c} \in \{0,1\}$ for every station--channel pair $(s,c) \in S\times C$, representing the proposition that station $s$ is assigned to channel $c$.  We encode the packing problem as a mixed integer program (MIP) with the objective of maximizing the value of on-air stations.

\begin{alignat}{2}
\text{maximize }   & \sum_{s \in \RepackingStations} \sum_{c \in D\left(s\right)} x_{s,c} \cdot \mathcal{V}\left(s\right)\ \nonumber \\ 
\text{subject to } &  x_{s,c} + x_{s',c'} \leq 1\ & \forall \{\left(s,c\right),\left(s',c'\right)\} \in I \label{MIP:interference}\\
& \sum_{c \in D\left(s\right)} x_{s,c} \leq 1\ & \forall s \in S \label{MIP:participants}\\
& \sum_{c \in D\left(s\right)} x_{s,c} = 1\ & \forall s \in \NonParticipatingStations \label{MIP:nonparticipants}\\
& x_{s,c} \in \left\{0,1\right\}  & \forall c \in D\left(s\right) \forall\ s \in S \label{MIP:integrality}
\end{alignat}

The objective function expresses that we want to maximize the sum of values of participating stations that remain on air. Constraint~\eqref{MIP:interference} prohibits infeasible channel assignments, Constraint~\eqref{MIP:participants} says that every station is either packed on a single channel or else is not packed, and Constraint~\eqref{MIP:nonparticipants} says that we must find a channel for all non-participating stations. Lastly, Constraint~\eqref{MIP:integrality} requires that each station's channel indicator variables be integral. Solving the MIP will yield the efficient allocation. 

The VCG price for a winning station $s$ is calculated as the difference (excluding the value of $s$) in the value of the optimal packing $\gamma^*$ and the value of the optimal packing $\gamma^*_{-s}$ when $s$ is added to $\NonParticipatingStations$. The price paid to any losing station is zero. The $|\Winners{}|$ pricing problems can be solved in parallel once the initial allocation has been found.

\section{The Reverse Auction}\label{sec:ReverseAuction}

We begin by giving an overview of our reverse auction simulator, followed by details about two key ingredients which we vary in our experiments: the scoring rule and the feasibility checker. We note that a major limitation of our simulator is that, in order to bypass some of the rules complexity of the reverse auction, we do not allow stations to bid on \VHF{} bands.

\subsection{Clock Auction Simulator}

Our clock auction simulator begins by computing opening prices for each station according to a scoring rule (described in Section~\ref{subsec:ScoringRule}). It then identifies as auction participants the set of stations for which the value is less than the opening price. We maintain an assignment of packed stations $\gamma$ throughout the auction, initially consisting of the non-participating stations. The auction then proceeds over a series of rounds, which consist of: (1) decrementing the clock and offering new prices, (2) collecting bids, and (3) processing bids. Only the processing step is not straightforward: Bids are considered sequentially, ordered by their price reduction in the current round with ties broken by a random seed. When processing a bid, the feasibility checker first determines whether a station is still packable along with the exited stations. If it is packable, the station either accepts the new clock price or exits the auction. Otherwise, the station becomes \emph{frozen}, meaning that it can no longer be packed in its original band. The station is now a \emph{winner} and will be paid according to its most recently accepted offer.


\subsection{Computing Prices}\label{subsec:ScoringRule}

The price $P$ for a station $s$ at round $t$ is computed by multiplying its volume with a base clock price $c_t$. The base clock price is decremented each round by $d_t$, the maximum of 5\% of its previous value or 1\% of its initial value. More formally,
\begin{gather}
d_t = \max \left\lbrace 0.05 \cdot c_{t-1}, 0.01 \cdot c_0 \right\rbrace \label{eq:decrement}\\
c_{t} = c_{t-1} - d_t\\
P_{s,t} = c_t \cdot \text{Volume}\left(s\right),
\end{gather}
where $c_0$ is the initial base clock price, used to compute the opening prices. A station's volume is set according to a \emph{scoring rule}.

\paragraph{FCC Scoring Rule}
The FCC uses the following scoring rule:
\begin{equation}
\text{Volume}\left(s\right)  = A \cdot \sqrt{\text{Interference}\left(s\right)} \cdot \sqrt{\text{Population}\left(s\right)},
\end{equation}
where Population$(s)$ is the interference-free population that $s$ reaches, Interference$(s)$ is based on the number of interference constraints $s$ participates in, and $A$ is a scaling constant used to make the maximum volume one million (for more details see Appendix D of the FCC's December 2014 public notice \cite{fcdec2014docket}). Volumes for each station are computed only once before the auction begins---they do not vary throughout the auction.
\paragraph{Unscored}
We consider an unscored alternative to the FCC scoring rule to determine the scoring rule's impact. Conceptually, this is the same as a scoring rule that assigns a volume of 1 to each station. In this case, the reverse auction can be understood as a greedy algorithm for packing stations: stations will exit exactly in descending order of value.

\subsection{Feasibility Checker}
The reverse auction must ask, hundreds of thousands of times per auction, whether or not a set of stations can be packed into a set of channels without causing interference. Since this is an NP-complete problem, and since the auction only has a finite amount of time to wait for each solution, there is no guarantee that a proof of feasibility or infeasibility will be found in the alloted time. If the feasibility checker times out, then the auction proceeds as if the problem was found to be infeasible. The feasibility checker impacts both efficiency and cost: if the feasibility checker can figure out how to squeeze more stations into the reduced \UHF{} band, then the auction will be more efficient; if the feasibility checker times out on a problem that really was feasible, then the auction either acquires a station that it did not want or pays a higher price than necessary to buy that station.  

More formally, the \emph{feasibility checking problem} is the task of determining whether or not there exists a packing $\gamma : S \to \overline{C}$ that assigns each station a channel from its domain that satisfies the interference constraints. A problem instance corresponds to a set of stations and a set of channels in which to pack them. Since the auction considers the feasibility of packing each active station one by one, for a given packing problem involving activate station $s^+$, we always have an assignment $\gamma^-$ that is a feasible assignment for the exited stations.

\paragraph{Greedy Feasibility Checker}
The \emph{greedy feasibility checker} is perhaps the simplest reasonable feasibility checker. It simply iterates through the domain of $s^+$ checking to see if any channel results in a feasible assignment when augmented with $\gamma^-$. It does not alter $\gamma^-$ in any way. If it cannot find a feasible channel for $s^+$, the greedy feasibility checker reports a timeout.

\paragraph{SATFC} 
\emph{SATFC}\cite{satfc_paper} is the feasibility checker adopted by the FCC for the incentive auction. It was tailor made for the incentive auction and the result of several years of research.  SATFC decomposes a problem instance into a SAT formula and uses a parallel portfolio of SAT solvers to tackle each problem. Unlike the greedy checker, SATFC is capable of proving that a given problem does not admit any feasible solution.

\section{Experiments and Results}\label{sec:Results}

\subsection{Station Information}
The interference constraints, station domains, and volumes applicable to the FCC's Incentive Auction are publicly available on the FCC Learn website \cite{stationinfofiles}. We used the most recent versions posted at the time of writing (from November 12, 2015). We restricted our attention to mainland US and Hawaii. We randomly sampled bidding station valuations from a publicly available model \cite{doraszelski2016ownership} using parameters obtained directly from its authors.\footnote{We excluded from our simulations 25 stations for which the authors of the model were unable to supply us with parameters: facility IDs 259, 4353, 8500, 14315, 17830, 27501, 34341, 34342, 34894, 38437, 38562, 41375, 51656, 57456, 57905, 66549, 68406, 69753, 70414, 70415, 70423, 70426, 70428, 71425, 168094.} This left us with 1399 US \UHF{} stations eligible to participate.\footnote{The FCC Learn website lists stations for which ``the auction system has determined that this station will always have a feasible channel assignment in its pre-auction band at all of the possible auction clearing targets.'' We ignored such stations in our simulations.} 

\begin{figure}[tp]
	\centering
	\includegraphics[width=.6\columnwidth]{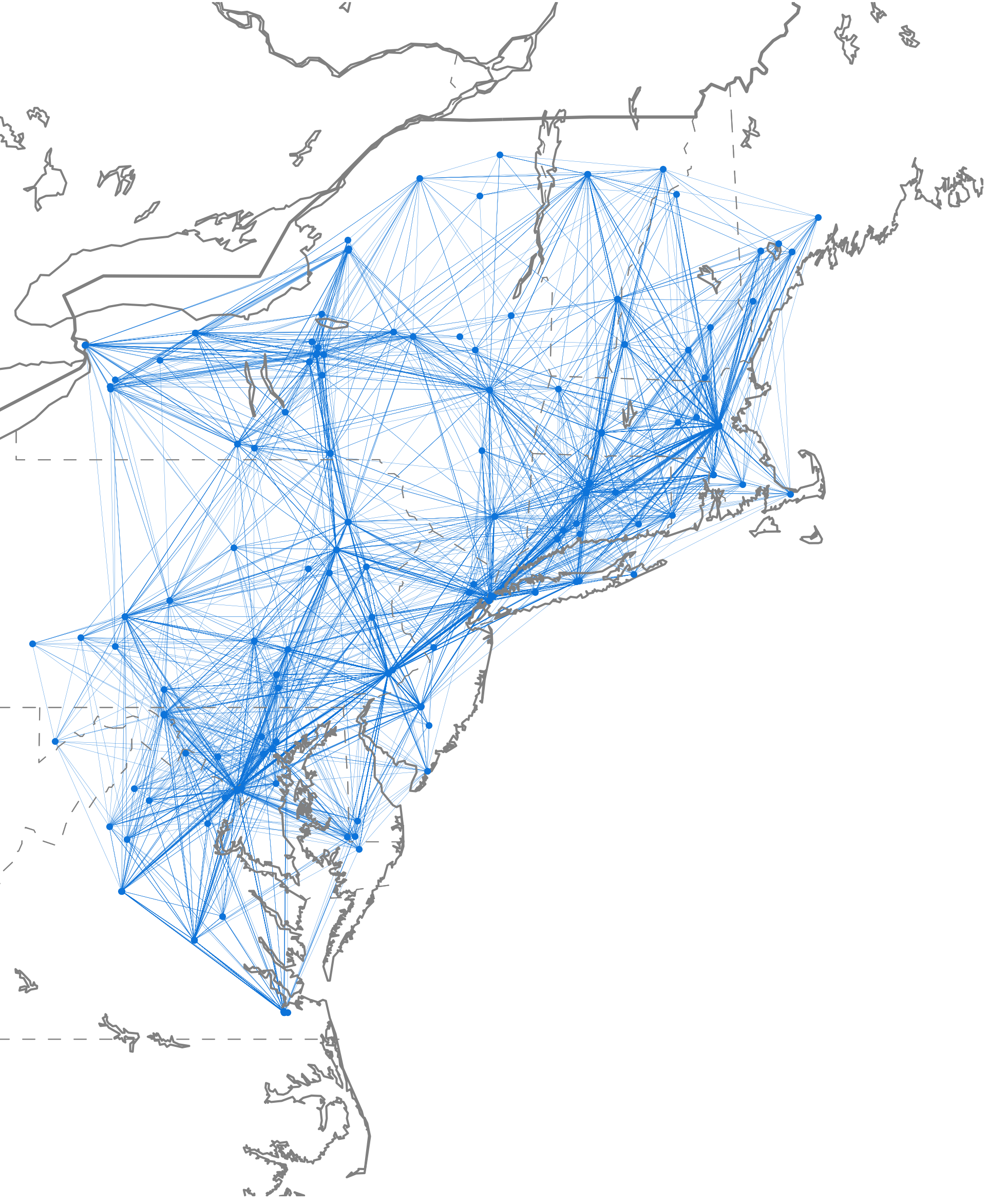}
	\caption{Interference graph of the set of 218 \UHF{} stations within two edges of a New York station. Each edge represents the existence of at least one pairwise binary constraint between two stations under a 126 MHz clearing target.}
	\label{fig:ConstraintGraph}
\end{figure}

We were unable to compute the VCG allocation with all of the eligible stations even after several days of computing time.\footnote{We did not work with approximate VCG allocations, as the estimation error on pricing can be quite dramatic. If the optimization error is bounded by $\varepsilon$ for each problem, then the cost will only be known to within $2|\Winners{}|\varepsilon$.} To make the problem tractable, we restricted ourselves in the following manner. Define the \emph{interference graph} $G$ as an undirected graph in which there is one vertex per station and an edge exists between two vertices $s$ and $s'$ if the corresponding stations participate together in any interference constraint: i.e., if there exist $c,c' \in \overline{C}$ such that $\{(s,c),(s',c')\} \in I$. We started with the set of stations in Manhattan, as this is one of the most densely connected areas of the interference graph. We then expanded our set of stations to include any station reachable by following at most two interference constraints from one of these initial stations. This gave us a set of 218 stations that we used for all of our experiments, ranging roughly between Buffalo, New York, Brunswick, Maine, and Suffolk, Virginia, and including Boston, Philadelphia and Washington DC. Figure~\ref{fig:ConstraintGraph} shows the interference graph restricted to our subset of stations.

In all of our examples we used a 126 MHz clearing target corresponding to $\overline{c} = 29$. This is the same clearing target that the FCC used in the first stage of the incentive auction. For the experiments using the FCC scoring rule we used an initial base clock price of \$900 and the FCC volumes. For the unscored experiments, we used an initial base clock price of \$900 million (corresponding to the maximum opening price offer in the reverse auction).

\subsection{Experiments}
First, we generated five value profiles. We then computed VCG allocations and prices for each value profile. We then ran SATFC + FCC scoring runs to mimic the actual auction, greedy feasibility checker + FCC scoring runs to investigate the impact of the feasibility checker, and SATFC unscored runs to investigate the impact of the scoring rule. All experiments were performed on a cluster environment where each node contained two Intel Xeon E5-2640 v2 processors (for a total of 16 cores per node) and 96 GB of RAM. Each node ran Red Hat Enterprise Linux Server release 6.7. We used CPLEX 12.6.2 and SATFC version 2.3.1. All MIPs were solved optimally to within $10^{-6}$ absolute MIP gap tolerance. 

\begin{figure}[t]
	\centering
	\includegraphics[width=\columnwidth]{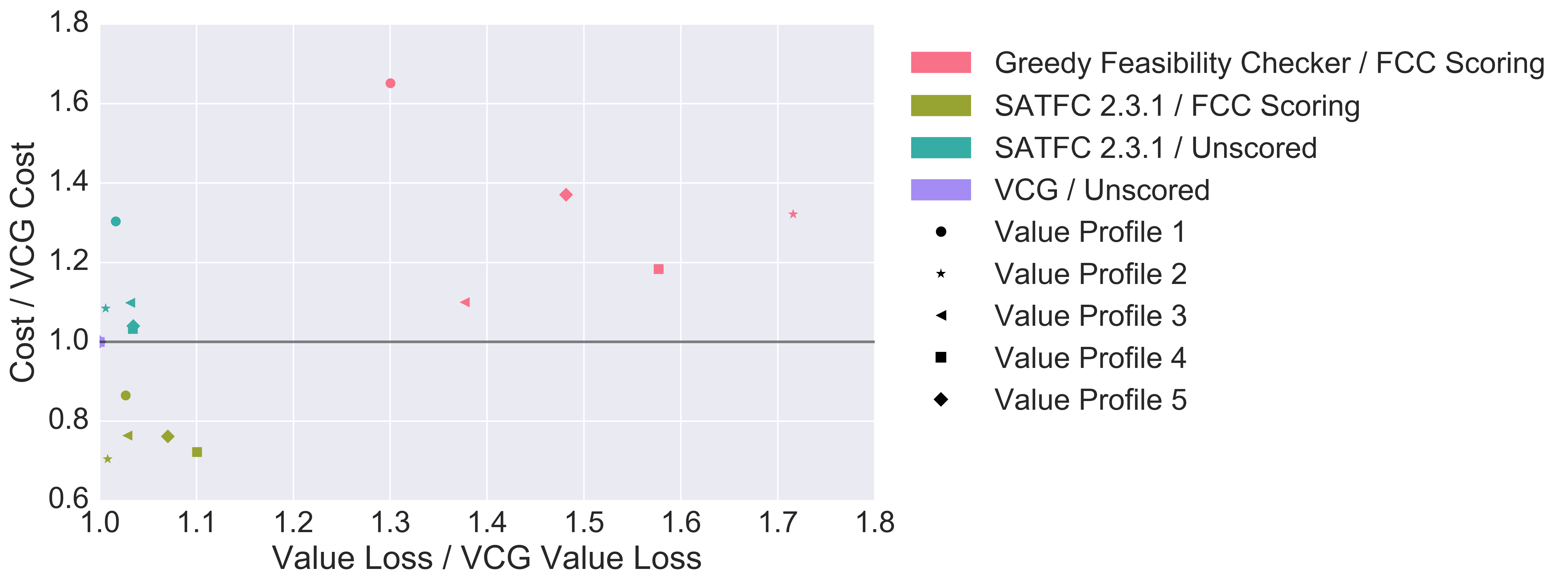}
	\caption{Fraction of VCG cost versus value loss ratio plotted for five different combinations of scoring rules and feasibility checkers (indicated by colors) for five different value profiles (indicated by markers). All VCG points lie at (1,1).}
	\label{fig:Simulations}
\end{figure} 

\subsection{Results}
Our results are summarized in Figure~\ref{fig:Simulations}, which plots the efficiency losses and costs of each simulation result relative to results from the VCG mechanism. We draw the following conclusions from these experiments. First, as the clustering of marks highlights, SATFC was a dramatic improvement over the naive feasibility checker according to both efficiency loss and cost. Averaging over all the observations, simulations with the naive feasibility checker cost 1.73 times more and lose 1.42 times as much broadcaster value. Second, the FCC scoring led to significant cost savings. On average over all observations, scored auctions cost $68.82\%$ of unscored auctions. Third, despite the clock auctions' much shorter run times when using SATFC, the efficiency performance of the unscored clock auction was very good. Averaging over all the unscored instances using SATFC, the value loss ratio of the clock auction was just 1.05. Finally, the FCC scoring did sometimes impose an efficiency penalty in exchange for the cost savings. Averaging over all the instances using SATFC, the value loss ratio was about $2.16\%$ greater using FCC scoring compared to using no scoring.   

\section{Conclusions and Future Work}

This report describes preliminary research into understanding tradeoffs and design decisions in the reverse auction portion of the FCC's inventive auction. We demonstrate that SATFC dramatically improves performance, that clock auctions can compute very good solutions with a tiny fraction of the computational effort, and that FCC scoring significantly reduces the cost of acquiring spectrum in the auction. Future follow up work might investigate a wider set of parameters of the auction or enhance the simulator to handle the \VHF{} band, or both. It also remains to be shown how robust our results are to changes in value profiles, participation assumptions, clearing targets, and larger subsets of stations.

\bibliographystyle{acm}
\bibliography{satfc}
										
\end{document}